\documentclass[12pt,preprint]{aastex}
\usepackage{epsfig}
\usepackage{natbib}
\usepackage{graphicx}
\usepackage{slashbox}
\usepackage{multirow}
\usepackage{lscape}
\usepackage{mathrsfs,amssymb}
\usepackage{amsmath}
\usepackage{subfigure}
\usepackage{amssymb}
\usepackage{multirow}
\usepackage{tabularx}
\usepackage{rotating}
\usepackage{amsmath}
\usepackage{ulem}
\usepackage{lineno}

\newcommand       \Angstrom     {\,{\rm \AA}}

\newcommand       \cm           {\,{\rm cm}}

\newcommand       \erg          {\,{\rm erg}}

\newcommand       \K            {\,{\rm K}}
\newcommand       \pc           {\,{\rm pc}}

\newcommand       \s            {\,{\rm s}}
\newcommand       \sr           {\,{\rm sr}}

\newcommand       \NH           {N_{\rm H}}

\newcommand       \simgt        {\gtrsim}

\newcommand       \mum          {\,{\rm \mu m}}
\newcommand       \ppm          {\,{\rm ppm}}
\newcommand       \Teff         {T_{\rm eff}}

\newcommand       \simali       {\sim\,}

\newcommand       \Aaro          {A_{3.3}}

\newcommand       \Acd           {A_{4.65}}
\newcommand       \ACD           {A_{4.4}}
\newcommand       \Adfa       {A_{6.85}}

\newcommand       \Adfb       {A_{7.25}}

\newcommand       \Acc        {A_{6.2}}

\newcommand \faliCH {f_{\rm aliCH}}
\newcommand \faliCD {f_{\rm aliCD}}
\newcommand \faroCD {f_{\rm aroCD}}
\newcommand \fD       {f_{\rm D}}
\newcommand       \NC         {N_{\rm C}}
\newcommand       \ND         {{N_{\rm D}}}

\newcommand       \NCaro      {N_{\rm C,aro}}
\newcommand       \NCali      {N_{\rm C,ali}}
\newcommand       \NHaro      {N_{\rm H,aro}}
\newcommand       \NHali      {N_{\rm H,ali}}
\newcommand       \NaroD         {{N_{\rm D,aro}}}
\newcommand       \NaliD         {{N_{\rm D,ali}}}

\newcommand  \IaliCDaroCHobs  {\left(I_{4.65}/I_{3.3}\right)_{\rm obs}}
\newcommand  \IaroCDaroCHobs {\left(I_{4.4}/I_{3.3}\right)_{\rm obs}}
\newcommand  \IaliCDaroCHmod {\left(I_{4.65}/I_{3.3}\right)_{\rm mod}}
\newcommand  \IaroCDaroCHmod {\left(I_{4.4}/I_{3.3}\right)_{\rm mod}}

\newcommand       \CabsPAH    {C^{\scriptscriptstyle\rm PAH}_{\rm abs}}
\countdef\decade=200
\advance\decade by \year
\countdef\hours=201
\advance\hours by \time
\divide\hours by 60
\countdef\mins=202
\advance\mins by \hours
\multiply\mins by 60
\multiply\hours by 100
\countdef\miltime=203
\advance\miltime by \hours
\advance\miltime by \time
\advance\miltime by -\mins
\def\today{\number\decade.\number\month.\number\day.\number\miltime}

\shorttitle{IR spectra of Deuterated PAHs}
\title{
Deuterated Polycyclic Aromatic Hydrocarbons
in the Interstellar Medium:
Constraints from the Orion Bar as Observed by the James Webb Space Telescope
\\{\small DRAFT: \today ~~}
}
\author{X.J.~Yang\altaffilmark{1,2}
            and Aigen Li\altaffilmark{2}
            }
\altaffiltext{1}{Department of Physics,
                      Xiangtan University,
                      411105 Xiangtan, Hunan Province, China;
                      {\sf xjyang@xtu.edu.cn}}
\altaffiltext{2} {Department of Physics and Astronomy,
                  University of Missouri,
                  Columbia, MO 65211, USA;
                  {\sf lia@missouri.edu}}

\begin{document}

\begin{abstract}
The gas-phase abundances of deuterium (D)
in the local interstellar medium (ISM) exhibit
considerable regional variations. Particularly,
in some regions the gas-phase D abundances are
substantially lower than the primordial D abundance
generated in the Big Bang, after subtracting
the astration reduction caused by the Galactic
chemical evolution. Deuterated polycyclic aromatic
hydrocarbon (PAH) molecules have been suggested
as a potential reservoir of the D atoms missing
from the gas-phase. Recent observations
from the James Webb Space Telescope's
Near Infrared Spectrograph
have revealed the widespread of deuterated
PAHs in the Orion Bar through their aliphatic
C--D emission at 4.65$\mum$ and possibly aromatic
C--D emission at 4.4$\mum$ as well.
To examine the viability of deuterated PAHs
as the D reservoir, we model the infrared (IR)
emission spectra of small PAH molecules
containing various aromatic and aliphatic D atoms
in the Orion Bar. We find that small deuterated
PAHs exhibit a noticeable emission band
at 4.4 or 4.65$\mum$ even if they contain
only one aromatic or aliphatic D atom.
We derive $\NaliD/\NH\approx3.4\%$,
the deuteration degree of PAHs
measured as the number of aliphatic
D atoms (relative to H),
from the observed intensity ratios
of the 4.65$\mum$ band to the 3.3$\mum$
aromatic C--H band.
The deuteration degree for
aromatically-deuterated PAHs
is less certain as C--N stretch also
contributes to the observed emission
around 4.4$\mum$. If we attribute it
exclusively to aromatic C--D, we derive
an upper limit of $\approx$\,14\%
on the deuteration degree,
which is capable of accounting for an appreciable
fraction of the missing D budget.
\end{abstract}



\section{Introduction\label{sec:intro}}
Deuterium (D) was exclusively generated in the Big Bang.
As the Galaxy evolves, D is converted into $^3$He,
$^4$He and heavier elements in stellar interiors
(a process known as ``astration'').
The D/H abundance therefore gradually
decreases from its primordial abundance of
D/H\,$\approx$\,26 parts per million (ppm)
predicted by the standard Big Bang
Nucleosynthesis (BBN) model,
to a present-day abundance of
D/H\,$\simgt$\,22$\ppm$,
according to the Galactic chemical
evolution (GCE) model.

However, observations of the local interstellar medium
(ISM) have revealed that the gas-phase interstellar D/H
abundance is appreciably lower than predicted from
the BBN and GCE models and varies considerably from
one sight line to another by a factor of $\simali$3
(Draine 2006; Linsky et al.\ 2006).
Such a low gas-phase D/H abundance
and its regional variations are puzzling.
In such a small scale as the local ISM,
variations in astration are not
expected to be significant enough to
account for the D/H variations.

\begin{figure}[htb]
\center{
\includegraphics[scale=0.5,clip]{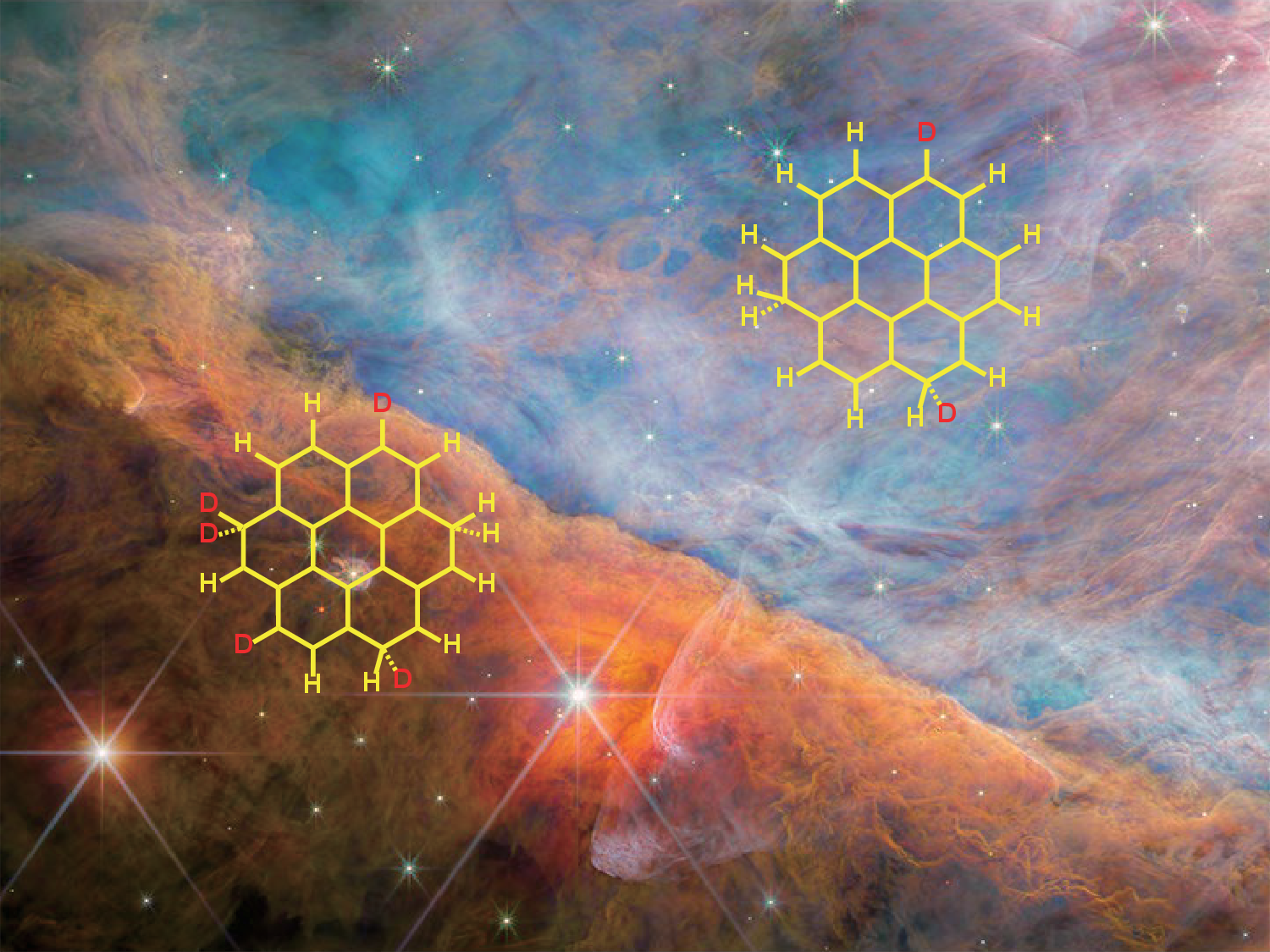}
}
\caption{\label{fig:OrionBar}\footnotesize
         Schematic illustration of
         the deuteration of PAHs
         in the Orion Bar.
         As discussed in Yang \& Li (2023),
         the incorporation of an D atom
         into a PAH molecule can occur
         either by replacing a peripheral H atom
         with an D atom or by inserting the (extra)
         D atom to an edge C atom so that one H atom
         and one D atom share one C atom.
         The former generates an aromatic
         C--D bond and the latter generates
         both an aliphatic C--H bond and
         an aliphatic C--D bond. Aliphatic C--D
         bonds can also be generated
         when two D atoms share one C atom.
         The background image of the Orion Bar
         was taken by the {\it Near Infrared Camera}
         (NIRCam) on JWST (Habart et al.\ 2024).
         }
\end{figure}

Where have all the missing D atoms gone?
One possible agent is polycyclic aromatic
hydrocarbon (PAH): all or part of the D atoms
missing from the gas-phase could have been
locked up in PAH molecules (Draine 2006).
When D atoms incorporated with PAHs,
both aromatic and aliphatic C--D bonds
could be formed (Yang \& Li 2023;
see Figure~\ref{fig:OrionBar}).
In principle, they could be revealed
by the aromatic and aliphatic C--D stretches
at $\simali$4.4 and 4.65$\mum$
(Hudgins et al.\ 2004; Yang \& Li 2023).
Indeed, in the past decades, such features
have been detected by the {\it Infrared Space Observatory}
and AKARI in several objects (Peeters et al.\ 2004;
Onaka et al. 2014; Doney et al. 2016).
Based on these detections,
Yang et al.\ (2020, 2021) and Yang \& Li (2023)
have determined the amount of D/H depleted
in PAHs and found that PAHs with aliphatic
C--D units could have tied up a substantial
amount of D/H and marginally account for
the missing D.

However, the low intensities of
the 4.4 and 4.65$\mum$ C--D emission bands
put them at the limit of ISO and AKARI.
This has changed with the launch and operation
of the {\it James Webb Space Telescope} (JWST).
Due to its unprecedented sensitivity,
JWST is expected to place their detection
on firm ground and enable far more detailed
analysis than previously possible.
Indeed, using the {\it Near Infrared Spectrograph}
(NIRSpec) on JWST, Peeters et al.\ (2024) performed
high-resolution near infrared  (IR) integral field
spectroscopic observations of the Orion Bar
and clearly detected the aliphatic C--D emission
band at 4.65$\mum$ and possibly the aromatic
C--D emission band at 4.4$\mum$ as well.
However, the detection of the 4.4$\mum$
band is less certain. The emission complex
around 4.4$\mum$ actually peaks at
$\simali$4.35$\mum$, which overlaps
with or is even dominated by C--N stretches.
The 4.65$\mum$ aliphatic C--D band was also
detected by JWST/NIRSpec in the bright
photodissociation region (PDR) position
in M17, a nearby massive star-forming
cloud (Boersma et al.\ 2023).
More recently, Draine et al.\ (2025) also reported
the detection of the 4.65$\mum$ aliphatic C--D
emission in four massive star-forming
regions in M51, the Whirlpool galaxy.
However, neither M17 nor M51 shows
appreciable aromatic C--D emission
at 4.4$\mum$.

In this work, we shall base on the JWST/NIRSpec
data of the Orion Bar to determine the depletion
of D atoms in PAHs. To this end, in \S\ref{sec:OrionBar}
we first summarize the JWST/NIRSpec detections
of the 4.4 and 4.65$\mum$ C--D emission bands
of the Orion Bar. We then model in \S\ref{sec:model}
the vibrational excitation of PAHs containing D atoms
and compute their IR emission spectra.
We consider three representative deuterated
PAH molecules with various aromatic
and aliphatic D atoms incorporated and derive
the power emitted from the 4.4 and 4.65$\mum$
bands ($I_{4.4}$ and $I_{4.65}$) as well as that from
the aromatic C--H stretch at 3.3$\mum$ ($I_{3.3}$).
We derive in \S\ref{sec:results} the degrees of
deuteration of PAHs by comparing the observed
band ratios $\IaroCDaroCHobs$ and $\IaliCDaroCHobs$
with the model band ratios. Finally, we summarize
our major results in \S\ref{sec:summary}.

\section{JWST Observations of the Orion Bar
           \label{sec:OrionBar}}
Lying in the Orion Nebula,
a prototypical star-forming region located
at a distance of $\simali$414$\pm$7$\pc$
(Menten et al.\ 2007), the Orion Bar refers to
the elongated rim near the ionization front
that separates the neutral cloud from
the ionized gas (see Figure~\ref{fig:OrionBar}).
It is illuminated by the O7V-type star
${\rm \theta}^1$ Ori C
(with an effective temperature of
$\Teff$\,$\approx$\,39,000$\pm$1,000$\K$),
the most massive member
of the Trapezium young stellar cluster,
which lies at the heart of the Orion Nebula
and $\simali$2$^{\prime}$ north-east of
the Bar (O'Dell 2001).
%

Recent JWST observations have enabled the detection
in the Orion Bar of a rich family of PAH emission bands
and an improved characterization of their detailed
spectral profiles and sub-components,
including the major bands
at 3.3, 6.2, 7.7, 8.6, 11.3, and 12.7$\mum$
and a wealth of weaker bands
(Chown et al.\ 2024; van De Putte et al.\ 2024;
Peeters et al.\ 2024).
As noticed previously from ground-based observations
(e.g., see Geballe et al.\ 1989; Sloan et al.\ 1997),
the 3.3$\mum$ band is accompanied by weaker bands
at 3.395, 3.403, 3.424, 3.46, 3.52, and 3.56$\mum$
which are perched on the top of a broad plateau.
These sub-features, collectively called
the ``aliphatic 3.4$\mum$ C--H'' stretching band,
demonstrate that the PAH molecules in the Orion Bar
have an aliphatic content (see Yang et al.\ 2017).

\begin{figure*}
\vspace{-8mm}
\centering{
\includegraphics[scale=0.75,clip]{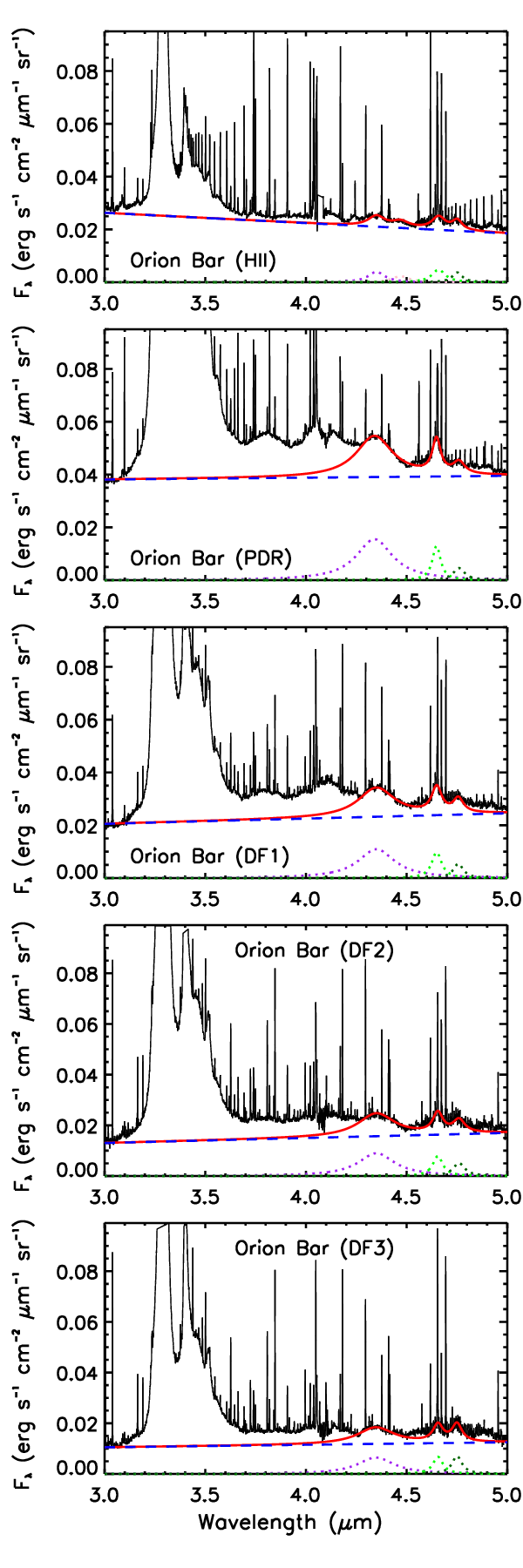}
}
\vspace{-5mm}
\caption{\footnotesize
  \label{fig:OrionFit1}
  JWST/NIRSpec spectra of the Orion Bar
  H\,{\sc ii} region, atomic PDR, DF1, DF2
  and DF3 regions in the 3--5$\mum$
  wavelength range. Also shown are our
  fitting results (solid red line)
  with several Drude profiles
  for the ``4.4'' (dotted purple line),
  4.65 (dotted green line) and
  4.75$\mum$ (dotted olive line) emission bands
  in combination with a linear continuum (dashed blue line).
         }
\end{figure*}

We show in Figure~\ref{fig:OrionFit1}
the 3--5$\mum$ JWST/NIRSpec spectra of
five representative regions in the Orion Bar:
H\,{\sc ii} region, atomic PDR,
and three ``dissociation front'' (DF) regions
(DF1, DF2 and DF3).
It is apparent that the Orion Bar shows
an asymmetric band (with a red wing)
centered at 4.646$\mum$ and potentially
an asymmetric, weaker band
(also with a red wing) centered
at 4.746$\mum$,\footnote{%
  We note that these bands have already
  been observed in the Orion Bar and
  several other H\,{\sc ii} regions by ISO and AKARI
  (Peeters et al.\ 2004; Onaka et al.\ 2014, 2022;
  Doney et al.\ 2016), although the band profiles
  could not be resolved due to the low angular
  and spectral resolution.
  }
which fall within the wavelength range of
the aliphatic C--D stretching mode of deuterated PAHs
(Hudgins et al.\ 2004; Buragohain et al.\ 2015;
Yang et al.\ 2020, 2021; Allamandola et al.\ 2021).
While the aromatic C--D stretch occurs
at $\simali$4.4$\mum$, the aliphatic C--D stretch
occurs near 4.6$\mum$
for super-deutereated PAHs\footnote{%
  By ``super-deutereated PAHs'' we mean that
  two D atoms or one D atom and one H atom
  share an C atom in a PAH molecule.
 In regions rich in H and D atoms, PAHs could
 be superhydrogenated and even super-deutereated
 by inserting an extra H or D atom to an edge C atom
 so that two H atoms or one H atom and one D atom
 or even two D atoms share one C atom.
 This converts aromatic C--H and C--D stretches
 to aliphatic (see Yang \& Li 2023).
 }
and near 4.7$\mum$ (2130$\cm^{-1}$)
for PAHs containing deuterated (i.e., D-substituted)
methyl side-groups (see Hudgins et al.\ 2004;
Yang \& Li 2023).

The JWST/NIRSpec spectra of the Orion Bar,
particularly the atomic PDR and DF1,
also display a broad, weak emission band
centered near 4.35$\mum$
(see Figure~\ref{fig:OrionFit1}).
For convenience,
in the following this band will be
interchangeably referred to
as the ``4.4$\mum$'' band
or the 4.35$\mum$ band.
We note that, although this band
occurs at a wavelength which is somewhat
shorter than the nominal wavelength of
4.4$\mum$ of the aromatic C--D stretch
of deuterated PAHs
in which a peripheral H atom
is replaced by an D atom,
it falls within the wavelength range of
$\simali$4.3--4.5$\mum$ predicted
from quantum chemical computations
of deuterated PAHs (Hudgins et al.\ 2004;
Yang et al.\ 2020, 2021).
However, as Peeters et al.\ (2024) had
already pointed out, this band also
coincides with a nitrile (--CN) stretch
at 4.38$\mum$ (2280$\cm^{-1}$;
Allamandola et al.\ 2021).\footnote{%
  It is interesting to note that the CN stretch
  of cyanonaphthalene, which consists of
  two fused benzene rings and substitutes a nitrile
  group for a hydrogen atom occurs
  at a much longer wavelength of
  $\simali$4.69$\mum$ (see Li et al.\ 2024).
  }
It is not clear how much of the power emitted
in the JWST-observed ``4.4$\mum$'' band
arises from aromatic C--D and how much
from C--N stretch. By exclusively attributing
the ``4.4$\mum$'' emission to aromatic C--D,
we will derive an upper limit on the amount of
aromatic D in PAHs.
%

To determine the amount of power emitted
in the aromatic and aliphatic C--D bands,
we fit the JWST/NIRSpec spectra of the five
regions of the Orion Bar
in the 3--5$\mum$ wavelength range
with three Drude profiles
at $\simali$4.35, 4.65 and 4.75$\mum$,
in combination with an underlying linear
continuum.\footnote{%
  For the H\,{\sc ii} region, two Drude components
  are required to account for the emission complex
  around 4.35$\mum$.
  }
These Drude profiles characterize
the aromatic and aliphatic C--D stretches.
We show the fitting results
in Figure~\ref{fig:OrionFit1}
and tabulate the so-derived emission intensities
for all three bands ($I_{4.4}$, $I_{4.65}$,
and $I_{4.75}$ in units $\erg\s^{-1}\cm^{-2}\sr^{-1}$)
in Table~\ref{Table:BandInt}.
For reference, we also tabulate
in Table~\ref{Table:BandInt}
the emission intensity of the 3.3$\mum$
aromatic C--H stretch ($I_{3.3}$) measured
by Peeters et al.\ (2024).
As the 4.75$\mum$ band often accompanies
the 4.65$\mum$ band and is often much
weaker than the 4.65$\mum$ band,
in this work we take the sum of $I_{4.65}$
and $I_{4.75}$ as the emission intensity of
the aliphatic C--D stretch.
In the following, unless otherwise stated,
``$I_{4.65}$'' represents the total emission
intensity of the aliphatic C--D stretches,
that is, $I_{4.65}+I_{4.75}$.

%

\begin{figure*}
\centering{
\includegraphics[scale=0.5,clip]{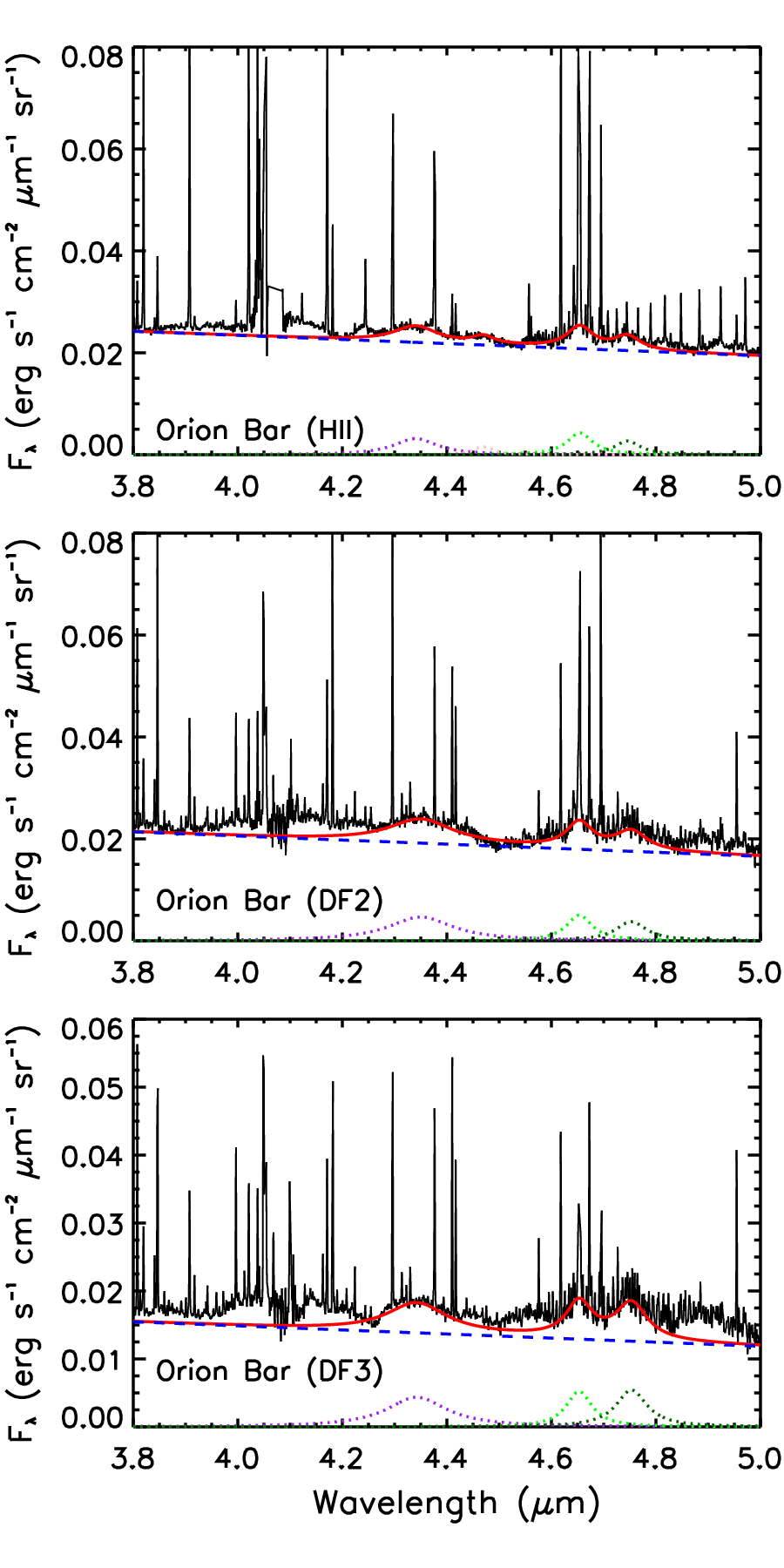}
}
\caption{\footnotesize
         \label{fig:OrionFit2}
         Same as Figure~\ref{fig:OrionFit1} but
         for the H\,{\sc ii} region, DF2 and DF3 regions
         in the 3.8--5$\mum$ wavelength range
         which allows us to define a more ``tight'' continuum.
         For the atomic PDR and DF1, we rely on
         the more ``generous'' continuum adopted
         in Figure~\ref{fig:OrionFit1}
         since it is hard to define a continuum
         in the 3.8--5$\mum$ wavelength range.
         }
\end{figure*}

Apparently, the derived band emission intensities
depend on the adopted underlying continuum.
The continua defined in Figure~\ref{fig:OrionFit1}
span the wavelength range of 3 and 5$\mum$
and are somewhat ``generous'', particularly for
the H\,{\sc ii}, DF2 and DF3 regions.
To assess the influences of the adopted continuum
on the band emission intensities,
we adopt  a``tight'' continuum defined
in a more narrow wavelength range
of 3.8--5.0$\mum$ and re-fit the
``4.4'', 4.65 and 4.75$\mum$ bands
for the H\,{\sc ii}, DF2 and DF3 regions.
No attempt is made to the atomic PDR
and DF1 since for these regions
it is hard to define a continuum
in the 3.8--5.0$\mum$ range.
In Figure~\ref{fig:OrionFit2} we show
the fitted spectra and in Table~\ref{Table:BandInt}
we tabulate the derived emission intensities
for the aromatic and aliphatic C--D bands,
which are somewhat smaller than that
when a ``generous'' continuum is adopted.

\section{Model Infrared Emission Spectra of
           Deuterated PAHs \label{sec:model}
         }
To facilitate the analysis of the IR emission bands of
deuterated PAHs in the Orion Bar as observed by JWST,
we set up a theoretical framework to model the IR emission
of PAHs containing both aromatic and aliphatic C--D bonds,
and relate the emission intensities of the 4.4$\mum$ aromatic
and 4.65$\mum$ aliphatic C--D bands (relative to
the 3.3$\mum$ aromatic C--H band)
to $\fD\equiv \ND/\left(\NH+\ND\right)$,
the degree of deuteration of a PAH molecule
measured as the number of D atoms ($\ND$)
relative to the sum of the number of H atoms
($\NH$) and $\ND$.

%
Lets consider a PAH molecule of
$\NCaro$ aromatic C atoms,
$\NHaro$ aromatic H atoms,
$\NHali$ aliphatic H atoms,
$\NaroD$ aromatic D atoms, and
$\NaliD$ aliphatic D atoms.
%
We assume that the aliphatic C--H and
C--D bonds are created by superhydrogenation
and superdeuteration
(i.e., an extra H or D atom is added
to an edge C atom so that two H atoms,
or one H atom and one D atom,
or two D atoms share one C atom;
see Figure~\ref{fig:OrionBar}).\footnote{%
  In the UV-rich Orion Bar, we believe that it is
  unlikely for PAHs to attain deuterated methyl
  side-groups.
  }

For such a molecule, we approximate
its absorption cross section by adding
five Drude functions to that of PAHs
of $\NCaro$ C atoms
and $\NHaro$ H atoms
which represent
the 3.4$\mum$ aliphatic C--H stretch,
the 6.85 and 7.25$\mum$ aliphatic
C--H deformation bending,
the 4.4$\mum$ aromatic C--D stretch,
and 4.65$\mum$ aliphatic C--D stretch:
\begin{eqnarray}
\nonumber
C_{\rm abs}(\NC;\lambda) & = & \CabsPAH(\NCaro,\NHaro;\lambda)\\
\nonumber
& + & \NHali \frac{2}{\pi}
    \frac{\gamma_{3.4} \lambda_{3.4} \sigma_{\rm int,3.3}
     \left(A_{3.4}/A_{3.3}\right)}
     {(\lambda/\lambda_{3.4}-\lambda_{3.4}/\lambda)^2
      +\gamma_{3.4}^2}\\
\nonumber
&+& \NHali \frac{2}{\pi}
     \frac{\gamma_{6.85} \lambda_{6.85}
     \sigma_{\rm int,6.2} \left(\Adfa/\Acc\right)}
     {(\lambda/\lambda_{6.85}-\lambda_{6.85}/\lambda)^2
    +\gamma_{6.85}^2}\\
\nonumber
&+& \NHali \frac{2}{\pi}
    \frac{\gamma_{7.25} \lambda_{7.25}
    \sigma_{\rm int,6.2} \left(\Adfb/\Acc\right)}
     {(\lambda/\lambda_{7.25}-\lambda_{7.25}/\lambda)^2
    +\gamma_{7.25}^2}\\
\nonumber
&+& \NaroD \frac{2}{\pi}
     \frac{\gamma_{4.4} \lambda_{4.4}
     \sigma_{\rm int,3.3} \left(\ACD/\Aaro\right)}
     {(\lambda/\lambda_{4.4}-\lambda_{3.3}/\lambda)^2
    +\gamma_{4.4}^2}\\
\label{eq:Cabs4}
&+& \NaliD \frac{2}{\pi}
    \frac{\gamma_{4.65} \lambda_{4.65}
    \sigma_{\rm int,3.3} \left(\Acd/\Aaro\right)}
     {(\lambda/\lambda_{4.65}-\lambda_{4.65}/\lambda)^2
    +\gamma_{4.65}^2} ~~,
\end{eqnarray}
where $\NC=\NCaro+\NCali$ is the total number
of C atoms contained in the molecule;
$\CabsPAH(\NCaro,\NHaro;\lambda)$
is the absorption cross section of
a pure aromatic astro-PAH molecule
of $\NCaro$ C atoms and $\NHaro$ H atoms
computed from the formulation of
Li \& Draine (2001), Draine \& Li (2007),
and Draine et al.\ (2021);
$\lambda_{3.4}=3.4\mum$ is the central
wavelength of the 3.4$\mum$ aliphatic C--H band;
$\lambda_{6.85}=6.85\mum$
and $\lambda_{7.25}=7.25\mum$
are respectively the central wavelengths
of the 6.85 and 7.25$\mum$ aliphatic
C--H deformation bands;
$\lambda_{4.4}=4.4\mum$
and $\lambda_{4.65}=4.65\mum$
are respectively the central wavelengths of
the 4.4 and 4.65$\mum$ C--D bands;
$\gamma_{3.4}\lambda_{3.4}$,
$\gamma_{4.4}\lambda_{4.4}$,
$\gamma_{4.65}\lambda_{4.65}$
$\gamma_{6.85}\lambda_{6.85}$,
and $\gamma_{7.25}\lambda_{7.25}$
are respectively the full widths at half
maximum (FWHMs)
of the 3.4, 4.4, 4.65, 6.85 and 7.25$\mum$ bands
($\gamma_{3.4}$, $\gamma_{4.4}$,
$\gamma_{4.65}$, $\gamma_{6.85}$,
and $\gamma_{7.25}$ are dimentionless
parameters; see Draine \& Li 2007);
$A_{3.3}$ and $A_{3.4}$ are the intensities of the
aromatic and aliphatic C--H stretches, respectively;
$A_{6.2}$ is the intensity of
the 6.2$\mum$ C--C stretch;
$A_{6.85}$ and $A_{7.25}$ are the intensities of
the aliphatic C--H deformation bands;
$A_{4.4}$ and $A_{4.65}$ are the intensities of
the aromatic and aliphatic C--D stretches;
and $\sigma_{{\rm int},3.3}$
is the integrated strengths per (aromatic)
C atom of the 3.3$\mum$ aromatic C--H stretch
(see Draine \& Li 2007).
We take $A_{3.4}/A_{3.3}=1.76$,
$\Adfa/\Acc=5.0$,
$\Adfb/\Acc=0.5$,
$A_{4.4}/A_{3.3}=0.56$ and
$A_{4.65}/A_{3.3}=1.04$
as computed by Yang et al.\ (2016, 2020, 2021)
and Yang \& Li (2023)\footnote{%
  These intrinsic band-strength ratios
  give $A_{4.65}/A_{3.4}\approx0.59$
  and $A_{4.65}/A_{4.4}\approx1.86$.
  }
and consistent with experimental
measurements (see Mori et al.\ 2022).
%
We set the FWHMs of the 4.4 and 4.65$\mum$ bands
to that observed in the Orion Bar,
$\gamma_{4.4}\lambda_{4.4}
= \gamma_{4.65}\lambda_{4.65}
= 0.047\mum$ (Doney et al.\ 2016).
%
We note that the absorption cross
sections given in eq.\,\ref{eq:Cabs4}
are not the results of density functional
theory (DFT) calculations
for PAH molecules with specific structures,
instead, they are simply a function of
$\NC$, $\NH$ and $\ND$,
the numbers of C, H and D atoms.
The three molecules studied in this work
(i.e., C$_{24}$H$_{12}$, C$_{32}$H$_{16}$,
and C$_{48}$H$_{18}$; see below)
are only representatives of molecules
of the same numbers of C, H and D atoms,
and their chemical structures
do not matter in the calculations.
%

Due to their small heat contents,
PAHs are transiently heated in the ISM
by single stellar photons.
They will not attain an equilibrium temperature,
instead, they will experience temperature spikes
and undergo temperature fluctuations.
Following Draine \& Li (2001), we calculate
the temperature probability distribution functions
and emission spectra of deuterated PAHs.
%
%
%
Let $dP$ be the probability that the temperature
of the PAH molecule will be in $[T,T+dT]$.
The emissivity (in unit of $\erg\s^{-1}\cm^{-1}$)
of this molecule becomes
\begin{equation}\label{eq:jlambda}
j_\lambda(\NC) = \int C_{\rm abs}(\NC,\lambda)\,
            4\pi B_\lambda(T)\,\frac{dP}{dT}\,dT  ~.
\end{equation}
We consider three parent PAHs:
C$_{24}$H$_{12}$ (like coronene),
C$_{32}$H$_{16}$ (like diindenoperylene)
and C$_{48}$H$_{18}$
(like [4]-rhombene).
We select these relatively small molecules
because, upon the absorption of an UV photon
in the ISM, only small PAHs can be excited
to high enough temperatures to effectively
emit around $\simali$4$\mum$.

For each molecule,
we consider a range of degrees of deuteration,
from zero deuteration all the way up to $\fD=50\%$;
i.e., $\ND=0, 1, 2, ..., 6$ for C$_{24}$H$_{12}$,
$\ND=0, 1, 2, ..., 8$ for C$_{32}$H$_{16}$, and
$\ND=0, 1, 2, ..., 9$ for C$_{48}$H$_{18}$.
For a given $\ND$, we vary $\NaliD$ from 0 to
$\ND$ and $\NaroD=\ND-\NaliD$.
For simplicity, we set $\NHali$ to be 10\%
of $\NC$. This is generally consistent with
the strength of the 3.4$\mum$ aliphatic C--H
emission (relative to the 3.3$\mum$ aromatic
C--H emission seen in the Orion Bar;
see Yang \& Li 2023). In principle, we should
consider different $\NHali/\NH$ ratios for
different regions in the Orion Bar so as to
agree with the JWST observations of the ratios
of the intensities of the 3.4$\mum$ band
($I_{3.4}$) to that of the 3.3$\mum$ band ($I_{3.3}$).
However, $I_{3.4}$ is often much smaller than
$I_{3.3}$ and the exact $\NHali$ value does not
appreciably affect the overall absorption cross
section of a PAH molecule (see eq.\,\ref{eq:Cabs4})
and therefore will not appreciably affect its temperature
probability distribution function
(see eq.\,\ref{eq:jlambda}).
As a result, the exact value
of $\NHali$ will have little influence on the model
$I_{4.4}$ and $I_{4.65}$ intensities.

For the illuminating interstellar radiation field,
we take the stellar model atmospheric spectrum
of Kurucz (1979) of $\Teff=40,000\K$,
the effective temperature of ${\rm \theta}^1$ Ori C
which illuminates the Orion Bar.
We take the starlight intensity
to be $U=10^4$,\footnote{%
  This is consistent with quantitative
  estimates of the UV radiation field
  of the Orion Bar:
  $G_0$\,$\approx$\,2.2--7.1$\times10^4$
  in Habing units (Bern\'e et al. 2022; Habart et al. 2023;
  Peeters et al. 2024) and $U\approx1.08\,G_0$
  (see Draine \& Li 2001).
  As we are mainly interested in the band ratios
  $I_{4.4}/I_{3.3}$ and $I_{4.65}/I_{3.3}$,
  the knowledge of the exact starlight intensity
  is not critical. As shown in Yang et al.\ (2016),
  the model IR emission spectra (scaled by
  the starlight intensity $U$) are essentially
  independent of $U$, characteristic of
  single-photon heating.
  }
where $U$ is defined as
\begin{equation}
U \equiv \frac{\int_{1\mu {\rm m}}^{912{\rm \Angstrom}}
               4\pi J_\star(\lambda)\,d\lambda}
              {\int_{1\mu {\rm m}}^{912{\rm \Angstrom}}
               4\pi J_{\rm ISRF}(\lambda)\,d\lambda} ~~,
\end{equation}
where $J_\star(\lambda)$ is the intensity of starlight
in the Orion Bar, and $J_{\rm ISRF}(\lambda)$ is
the starlight intensity of the solar neighbourhood
interstellar radiation field (ISRF) of
Mathis, Mezger \& Panagia (1983; MMP83).

\section{Results and Discussion
           \label{sec:results}
           }
\begin{figure*}
\centering{
\includegraphics[scale=0.5,clip]{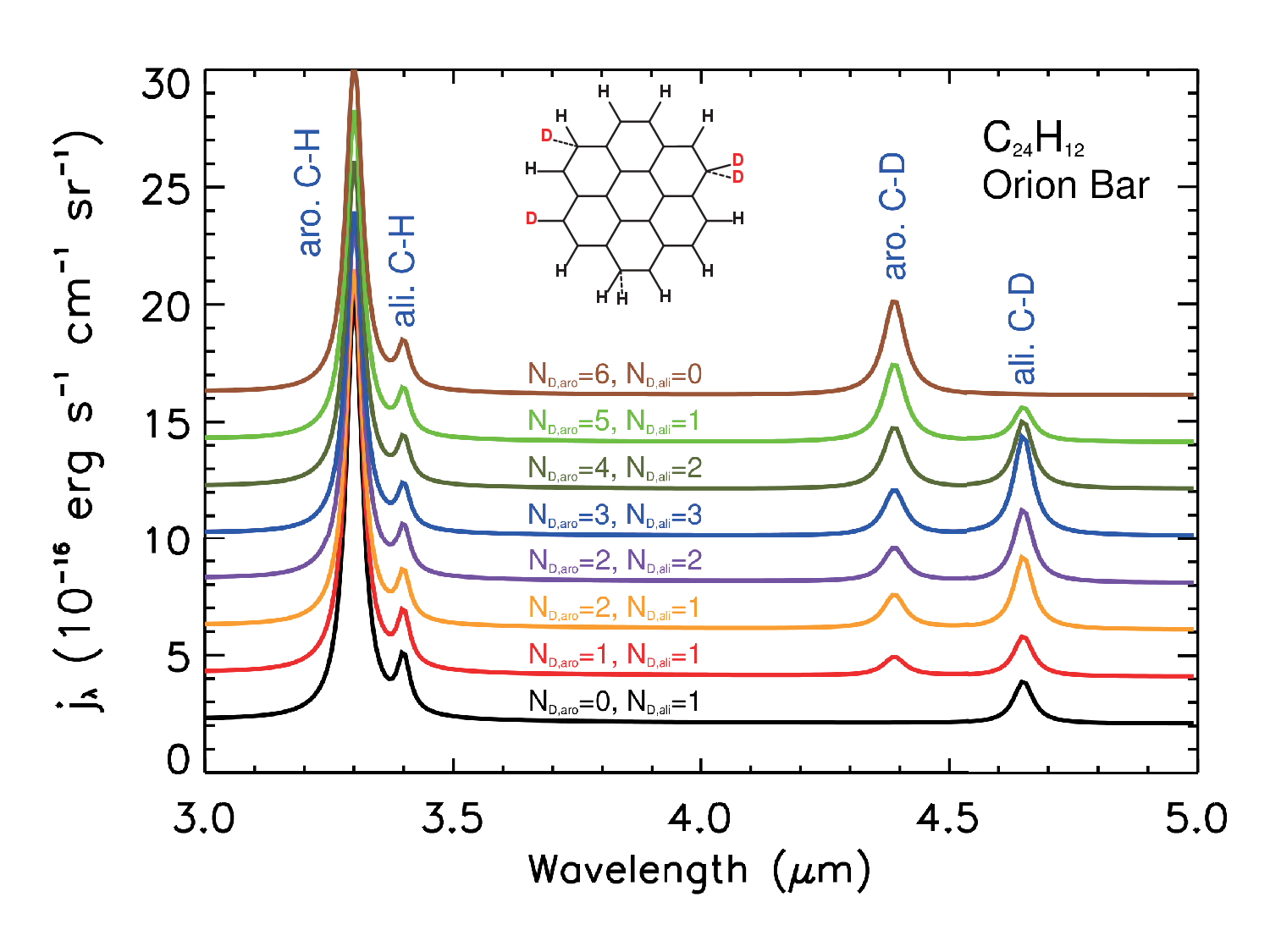}
}
\caption{\footnotesize
         \label{fig:C24H12spectra}
         Model IR emission spectra of deuterated
         C$_{24}$H$_{12}$ containing various
         aromatic and aliphatic D atoms
         in the Orion Bar.
         The inserted molecule structure
         is just for illustrative purpose
         and does not really specify
         the chemical structure of
         the molecule for which
         the IR emission spectra
         are calculated.
         }
\end{figure*}

\begin{figure*}
\centering{
\includegraphics[scale=0.5,clip]{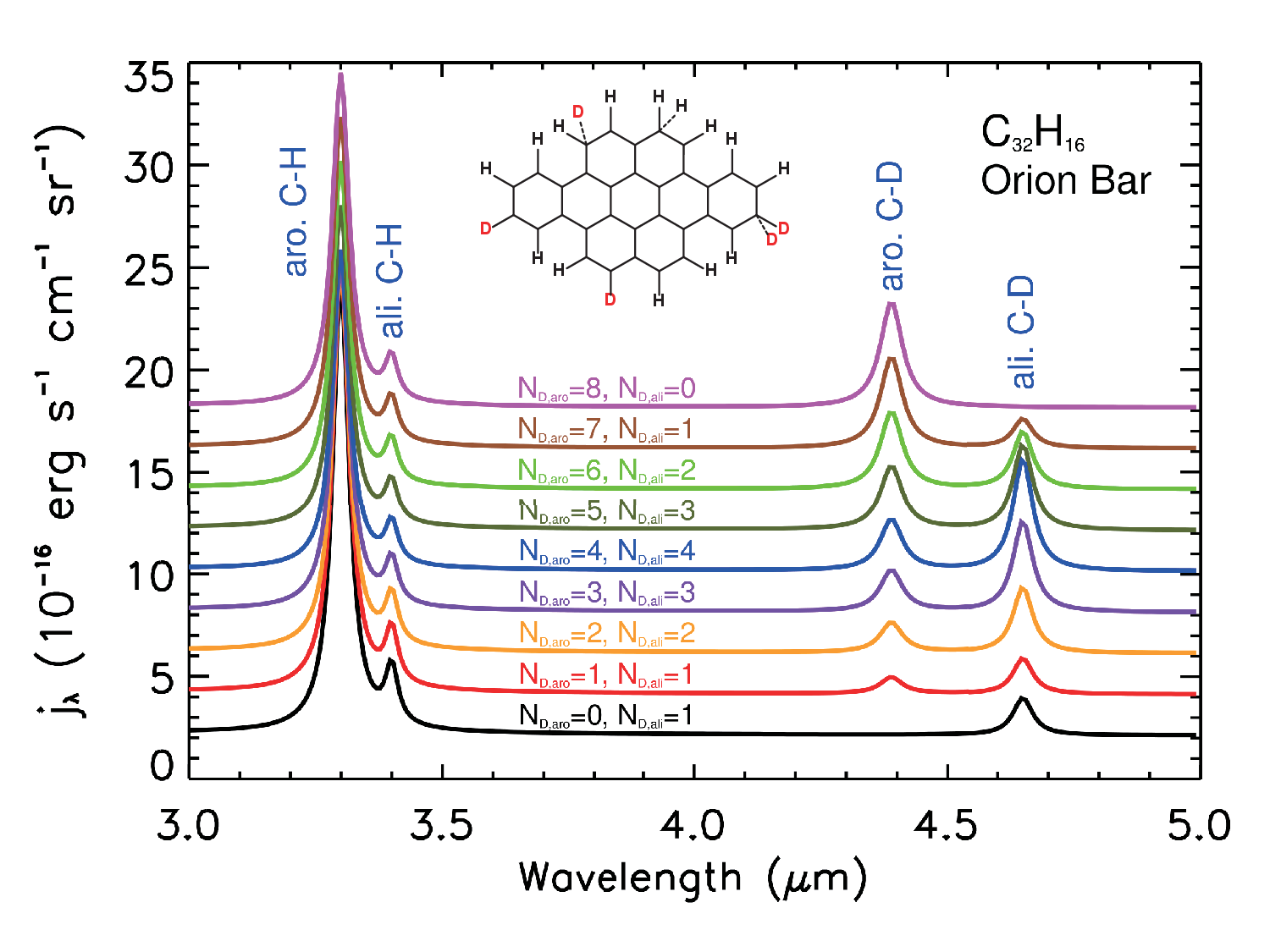}
}
\caption{\footnotesize
         \label{fig:C32H16spectra}
         Same as Figure~\ref{fig:C24H12spectra}
         but for C$_{36}$H$_{16}$.
         }
\end{figure*}

\begin{figure*}
\centering{
\includegraphics[scale=0.5,clip]{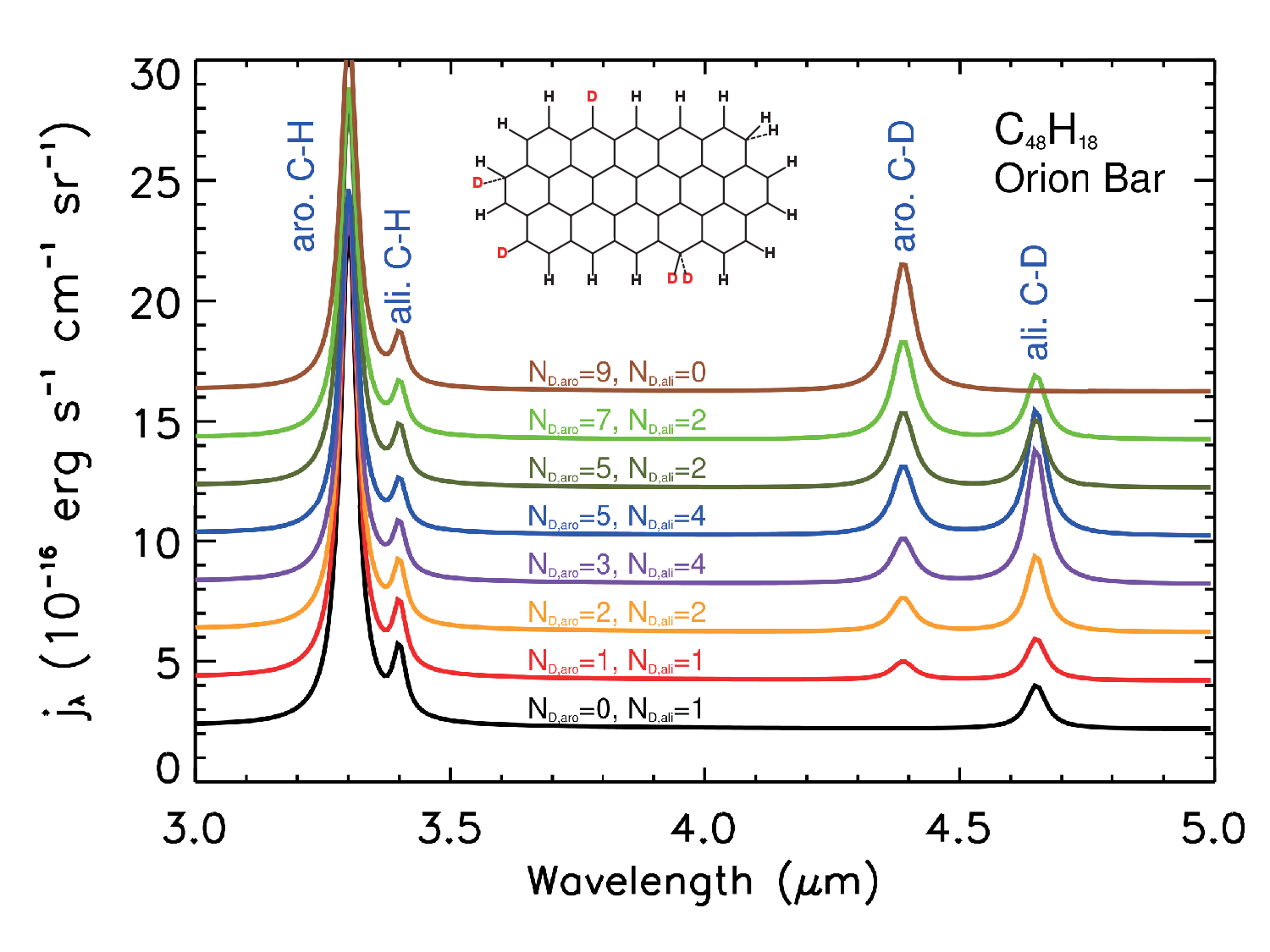}
}
\caption{\footnotesize
              \label{fig:C48H18spectra}
              Same as Figure~\ref{fig:C24H12spectra}
              but for C$_{48}$H$_{18}$.
              }
\end{figure*}

We calculate the IR emission spectra of
various deuterated species of three parent
molecules C$_{24}$H$_{12}$,
C$_{32}$H$_{16}$, and C$_{48}$H$_{18}$,
illuminated by ${\rm \theta}^1$ Ori C
in the Orion Bar. For illustration, we show in
Figures~\ref{fig:C24H12spectra}--\ref{fig:C48H18spectra}
the model emission spectra in the wavelength
range of 2.5--5.0$\mum$ for a number
of deuterated species of C$_{24}$H$_{12}$,
C$_{32}$H$_{16}$, and C$_{48}$H$_{18}$.
We see that the 4.4 and 4.65$\mum$ aromatic
and aliphatic C--D emission bands are clearly
visible even just with $\NaroD=1$ and $\NaliD=1$,
and become stronger as $\NaroD$ and $\NaliD$ increase.
For molecules with $\NaroD=\NaliD$,
the 4.65$\mum$ band is stronger than
the 4.4$\mum$ band by a factor of $\simali$2.
This is because the intrinsic band strength of
the 4.65$\mum$ aliphatic C--D stretch
is stronger than that of the 4.4$\mum$ aromatic
C--D stretch ($A_{4.65}/A_{4.4}\approx 1.86$;
see Yang \& Li 2023).
Also, as expected, the 3.4$\mum$ aliphatic
C--H emission remains essentially invariant
since the number of aliphatic C--H bonds is
fixed to be 10\% of $\NC$
(see \S\ref{sec:model}).

For each molecule with a given $\NaroD$
and $\NaliD$, we derive $\IaroCDaroCHmod$
and $\IaliCDaroCHmod$,
the model emission intensity ratios of
the 4.4 and 4.65$\mum$ bands
to the 3.3$\mum$ band.
They are calculated from
\begin{equation}\label{eq:Iratiomod}
\left(\frac{I_{4.4}}{I_{3.3}}\right)
= \frac{\int_{4.4}\Delta j_\lambda(\NC)\,d\lambda}
{\int_{3.3}\Delta j_\lambda(\NC)\,d\lambda} ~~,~~
\left(\frac{I_{4.65}}{I_{3.3}}\right)
= \frac{\int_{4.65}\Delta j_\lambda(\NC)\,d\lambda}
{\int_{3.3}\Delta j_\lambda(\NC)\,d\lambda} ~~,
\end{equation}
where $I_{3.3}$, $I_{4.4}$ and $I_{4.65}$ are
respectively the intensities of the 3.3$\mum$
aromatic C--H emission band,
the 4.4$\mum$ aromatic C--D band,
and the 4.65$\mum$ aliphatic C--D band;
and $\int_{3.3}\Delta j_\lambda(\NC)\,d\lambda$,
$\int_{4.4}\Delta j_\lambda(\NC)\,d\lambda$, and
$\int_{4.65}\Delta j_\lambda(\NC)\,d\lambda$
are respectively the band-integrated
excess emission of the 3.3, 4.4 and
4.65$\mum$ bands of deuterated PAHs.

\begin{figure*}
\centering{
\includegraphics[scale=0.55,clip]{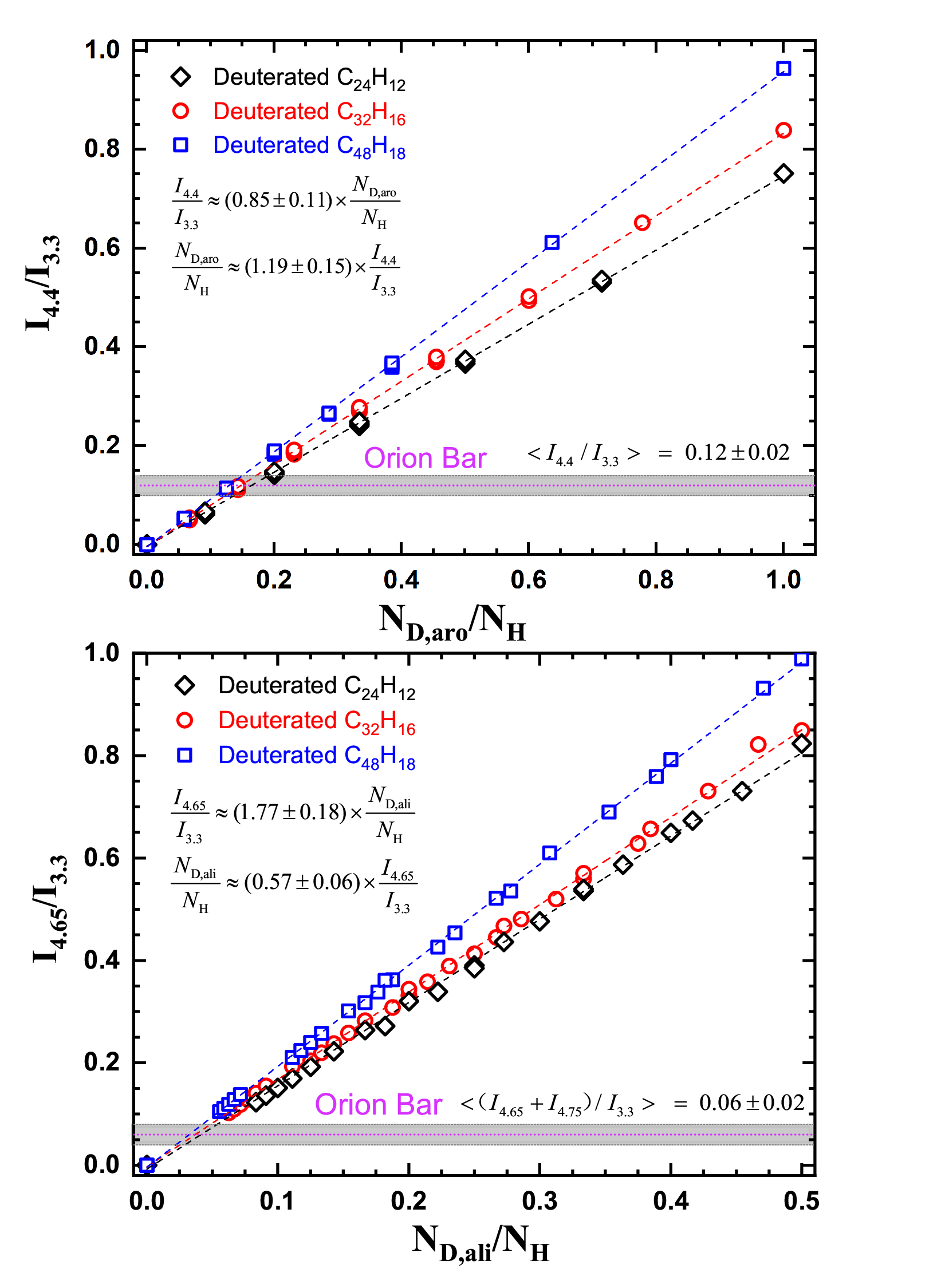}
}
\caption{\footnotesize
         \label{fig:Iratio_Nratio}
         Top panel (a): Model-calculated intensity ratios
         $\IaroCDaroCHmod$ as a function of $\NaroD/\NH$
         for deuterated C$_{24}$H$_{12}$,
         C$_{32}$H$_{16}$, and C$_{48}$H$_{18}$.
         The dotted horizontal line shows
         the JWST/NIRSpec-observed $\IaroCDaroCHobs$ ratio
         averaged over the five regions in the Orion Bar,
         while the shadow shows the standard deviation.
         Bottom panel (b): Same as (a) but for
         $\IaliCDaroCHmod$ as a function of $\NaliD/\NH$.
         }
\end{figure*}

To relate the model $\IaroCDaroCHmod$ and
$\IaliCDaroCHmod$ band-intensity ratios to
the degrees of deuteration, we show in
Figure~\ref{fig:Iratio_Nratio}
the band-intensity ratios
$\IaroCDaroCHmod$ and $\IaliCDaroCHmod$
as a function of $\NaroD/\NH$
and $\NaliD/\NH$, respectively.
The band-intensity ratios are computed
for the various deuterated species of
C$_{24}$H$_{12}$, C$_{32}$H$_{16}$,
and C$_{48}$H$_{18}$, containing various
aromatic and aliphatic D atoms.
It is apparent that for each parent molecule,
$\IaroCDaroCHmod$ linearly increases with
$\NaroD/\NH$ and $\IaliCDaroCHmod$ linearly
increases with $\NaliD/\NH$.
For larger PAHs, the increase is somewhat steeper
(i.e., the slopes
$d\IaroCDaroCHmod/d\left(\NaroD/\NH\right)$
and $d\IaliCDaroCHmod/d\left(\NaliD/\NH\right)$
are slightly higher).
This is because, a larger PAH molecule
has more degrees of freedom and a larger
heat capacity. Upon absorption of an UV stellar
photon, the molecule is excited to lower temperatures
and emits more effectively at longer wavelengths
(e.g., 4.4 and 4.65$\mum$) than at shorter wavelengths
(e.g., 3.3$\mum$). Therefore, for a given $\ND/\NH$,
larger $\IaroCDaroCHmod$ and $\IaliCDaroCHmod$
band-ratios are expected.
By averaging over all the deuterated species of
all three parent molecules, we obtain
\begin{equation}\label{eq:Iratio_Nratio}
\frac{\NaroD}{\NH} \approx
\left(1.19\pm0.15\right)\times\IaroCDaroCHmod ~~,~~
\frac{\NaliD}{\NH}\approx
\left(0.57\pm0.06\right)\times\IaliCDaroCHmod ~~.
\end{equation}
It is straightforward to apply eq.\,\ref{eq:Iratio_Nratio}
to observationally determined band-intensity ratios
$\IaroCDaroCHobs$ and $\IaliCDaroCHobs$
to derive the deuteration of PAHs.
%

As summarized in \S\ref{sec:OrionBar},
the JWST/NIRSpec observations of the Orion Bar
clearly detected the 4.65 and 4.75$\mum$
aliphatic C--D emission bands in the H\,{\sc ii}
region, atomic PDR, and three DF regions
(Peeters et al.\ 2024).
By adopting a ``generous'' continuum
(see Figure~\ref{fig:OrionFit1}),
the JWST-observed emission intensity ratios
of the 4.65$\mum$ aliphatic C--D stretch
to the 3.3$\mum$ aromatic C--H stretch
are $I_{4.65}/I_{3.3}$\,$\approx$\,0.057, 0.023,
0.030, 0.031 and 0.035 for the Orion Bar
H\,{\sc ii} region, atomic PDR, DF1, DF2
and DF3, respectively.
If we adopt a ``tight'' continuum
(see Figure~\ref{fig:OrionFit2}),
we derive $I_{4.65}/I_{3.3}$\,$\approx$\,0.038,
0.020, and 0.032 for the H\,{\scriptsize II} region,
DF2 and DF3, respectively
(see Table~\ref{Table:BandInt}).
Similarly, we determine the JWST-observed emission
intensity ratios of the 4.75$\mum$ aliphatic C--D
stretch to the 3.3$\mum$ aromatic C--H stretch
to be $I_{4.75}/I_{3.3}$\,$\approx$\,0.033, 0.011,
0.017, 0.023 and 0.043 for the H\,{\sc ii} region,
atomic PDR, DF1, DF2 and DF3, respectively.
With a ``tight'' continuum, we obtain
$I_{4.75}/I_{3.3}$\,$\approx$\,0.024, 0.018,
and 0.040 for the H\,{\scriptsize II} region,
DF2 and DF3, respectively.
As we aim at maximizing the amount of D
which PAHs could accommodate,
in the following we will confine us to
the ``generous'' continuum cases.

As mentioned in \S\ref{sec:OrionBar},
we shall take the sum of $I_{4.65}$
and $I_{4.75}$ as the emission intensity of
the aliphatic C--D stretch,
leading to $I_{``4.65''}/I_{3.3}$\,$\approx$\,0.090,
0.034, 0.047, 0.054 and 0.077 for the Orion Bar
H\,{\sc ii} region, atomic PDR, DF1, DF2
and DF3, respectively.
According to eq.\,\ref{eq:Iratio_Nratio},
these emission intensity ratios
translate into deuteration fractions of
$\NaliD/\NH$\,$\approx$\, 0.051, 0.019,
0.027, 0.031 and 0.044 for the H\,{\sc ii} region,
atomic PDR, DF1, DF2 and DF3, respectively.
By averaging over all five regions, we obtain
$\langle I_{``4.65''}/I_{3.3}\rangle$\,$\approx$\,0.06$\pm$0.02
and $\langle\NaliD/\NH\rangle$\,$\approx$\,3.4\%
for the Orion Bar. Such a deuteration fraction
is too small to account for the missing D in the ISM,
which requires $\ND/\NH$\,$\simali$20\%
(see Yang et al.\ 2020, 2021; Yang \& Li 2023).
It is likely that the bulk of the missing D may
have been locked up in large PAHs which do not
emit appreciably at 4.65$\mum$
(B.T.~Draine 2024, private communication).

As discussed in \S\ref{sec:OrionBar},
the case for the 4.4$\mum$ aromatic
C--D stretch is more complicated
since the observed emission complex
around 4.4$\mum$ is also contributed
or even dominated by C--N stretches.
If we attribute it exclusively to aromatic
C--D, an upper limit on the deuteration
fraction $\NaroD/\NH$ will be placed.
As tabulated in Table~\ref{Table:BandInt},
if we adopt a ``generous'' continuum
(see Figure~\ref{fig:OrionFit1}),
the observed emission intensity ratios
of the ``4.4$\mum$'' band
to the 3.3$\mum$ band
are $I_{``4.4''}/I_{3.3}$\,$\approx$\,0.093,
0.118, 0.124, 0.129 and 0.144
for the Orion Bar H\,{\sc ii} region,
atomic PDR, DF1, DF2 and DF3, respectively.
These emission intensity ratios
translate into deuteration fractions
of $\NaroD/\NH$\,$\approx$\,0.110, 0.141,
0.148, 0.153 and 0.171 for the H\,{\sc ii} region,
atomic PDR, DF1, DF2 and DF3, respectively.
By averaging over all five regions, we obtain
a mean intensity ratio of
$\langle I_{``4.4''}/I_{3.3}\rangle$\,$\approx$\,0.12$\pm$0.02
and $\langle\NaliD/\NH\rangle$\,$\approx$\,14\%
for the Orion Bar.
By adding $\NaliD$ to $\NaroD$,
we derive an upper limit of
$\ND/\NH$\,$\approx$\,17\%
for the deuteration degree of
small PAH molecules in the Orion Bar.
This is capable of accounting for
a large fraction of the missing D budget.
We should emphasize that these quantities
(i.e., $I_{``4.4''}/I_{3.3}$, $\NaliD/\NH$,
and $\ND/\NH$) are the upper limits.
%

We note that, as summarized in
Table~7 of Yang \& Li (2023),\footnote{%
  In Table~7 of Yang \& Li (2023),
  a per cent (``\%") symbol was missing
  in the numbers listed for
  $\left(I_{3.4}/I_{3.3}\right)_{\rm obs}$, $\faliCH$,
  $\left(I_{4.4}/I_{3.3}\right)_{\rm obs}$, $\faroCD$,
  $\left(I_{4.65}/I_{3.3}\right)_{\rm obs}$,
  and $\faliCD$. For example, $59.9\pm4.1$
  should be $59.9\%\pm4.1\%$
  for G75.78+0.34's
  $\left(I_{3.4}/I_{3.3}\right)_{\rm obs}$,
  and $3.3\pm1.3$ should be $3.3\%\pm1.3\%$
  for G75.78+0.34's $\faroCD$.
  }
ISO and AKARI have previously detected
the aromatic and aliphatic C--D emission
in about a dozen sources
(Verstraete et al.\ 1996; Peeters et al.\ 2004;
Onaka et al.\ 2014, 2022; Doney et al.\ 2016).
Based on these data, Yang \& Li (2023) found
that the PAH deuteration degrees of some sources
can be as high as $\simali$20\%,
implying that PAHs could be a plausible sink
of the missing D atoms.
Particularly, the aliphatic D/H derived from
the 4.65$\mum$ band of these sources is
much higher than that derived here for
the Orion Bar. Future JWST/NIRSpec observations
with high spectral resolution
and superb sensitivity will be valuable
for placing these detections on firmer ground
and for attesting the deuteration degree
derivation of Yang \& Li (2023).\footnote{%
    Boersma et al.\ (2023) performed
    JWST/NIRspec observations of
    the massive star-forming cloud M17
    and detected the 4.65$\mum$ aliphatic
    C--D emission band, but they found no
    hint of the aromatic C--D stretch
    between 4.36 and 4.43$\mum$.
    They derived from $I_{4.65}/I_{3.4}$---the
    emission-intensity ratio of the 4.65$\mum$
    band to the 3.4$\mum$ aliphatic C--H
    band---a degree of aliphatic deuteration of
$\ND/\left(\NH+\ND\right)$\,=\,$31\%\pm12.7\%$
for the bright-PDR position in M17.
However, this deuteration degree only
applies to the aliphatic C--H units,
which are only a small fraction of
the PAHs in the M17 PDR for which
$I_{3.4}/I_{3.3}\approx0.10$
(Boersma et al.\ 2023).
With $I_{4.65}/I_{3.3}\approx0.037$
(Boersma et al.\ 2023),
we derive $\NaliD/\NH\approx2.1\%$
for the M17 PDR, which is close to that
estimated by Yang \& Li (2023) based on
the AKARI data of Onaka et al.\ (2014).
}


Finally, we also note that,
following the absorption of an energetic photon,
a PAH molecule has three major competing decay
channels to relax its energy:
thermal emission (i.e., ``heat''),
photoionization, and photodissociation
(see Appendix A in Li \& Lunine 2003).
When we model the vibrational excitation
of deuterated PAHs in this work
(see \S\ref{sec:model}),
we assume that the energy of an absorbed photon
is fully converted to vibrational energy of the molecule,
i.e., we neglect the energy lost in the form of
photoelectrons (e.g., see Sidhu et al.\ 2022)
and bond dissociation
as well as possible fluorescent emission
of optical photons (e.g., see Li \& Draine 2002).
If we take into account the energy lost in these
channels, the overall IR emission will be smaller.
However, this will not affect the PAH deuteration
degree derivation since what we rely on are
the band-intensity {\it ratios}, not the absolute
power emitted from a C--D band.

\section{Summary}\label{sec:summary}
We have modeled the IR emission spectra of
deuterated PAH molecules in the Orion Bar.
We have considered various deuterated species
of three small PAH molecules
(C$_{24}$H$_{12}$,
C$_{32}$H$_{16}$, and C$_{48}$H$_{18}$)
which are the most effective emitters
in the C--H and C--D stretching bands.
These deuterated species contain a range
of aromatic and aliphatic C--D bonds and
emit at the 3.3$\mum$ aromatic C--H stretch,
4.4$\mum$ aromatic C--D stretch,
and 4.65$\mum$ aliphatic C--D stretch.
The major results are as follows:
\begin{enumerate}
\item Small deuterated PAHs exhibit a noticeable
          emission band at 4.4 or 4.65$\mum$
          even if they contain {\it only} one aromatic
          or aliphatic D atom. With the same amount
          of aromatic and aliphatic D atoms,
          the 4.65$\mum$ aliphatic C--D band
          is stronger than the 4.4$\mum$ aromatic
          C--D band by a factor of $\simali$2
          since the intrinsic strength of the 4.65$\mum$
          aliphatic C--D stretch is about twice strong
          as that of the 4.4$\mum$ aromatic C--D stretch.
\item     The fractions of aromatic and aliphatic D atoms
          in deuterated PAHs can be derived from
          the observed band-intensity ratios
          of the 4.4 and 4.65$\mum$ C--D
          emission bands to the 3.3$\mum$
          C--H band through
          $\left(\NaroD/\NH\right)\approx
\left(1.19\pm0.15\right)\times\IaroCDaroCHobs$
and $\NaliD/\NH\approx
\left(0.57\pm0.06\right)\times\IaliCDaroCHobs$.
\item     The JWST/NIRSpec observations of
          the Orion Bar reveal mean band-intensity
          ratios of $\IaroCDaroCHobs\approx0.12\pm0.02$
          and $\IaliCDaroCHobs\approx0.06\pm0.02$.
          These band-intensity ratios translate into
          $\NaroD/\NH\approx14\%$,
          $\NaliD/\NH\approx3.4\%$,
          and a total deuteration fraction of
          $\ND/\NH\approx17\%$.
          However, such a deuteration fraction
          should be considered as an upper limit
          since the JWST-observed 4.4$\mum$ emission
          complex is also contributed or even dominated
          by C--N stretches, and the adopted continuum
          underneath this band is somewhat generous
          and could overestimate the band flux.
          If the observed 4.4$\mum$ band indeed arises
          exclusively from aromatic
          C--D stretch, deuterated PAHs are able to
          account for a large fraction of the missing
          D budget in the ISM.
\end{enumerate}

\acknowledgments{%
We thank B.T.~Draine, A.N.~Witt
and the anonymous referee
for valuable suggestions.
We thank E.~Peeters for providing us
the JWST/NIRSpec spectra of the Orion Bar.
XJY is supported in part by
NSFC~12333005 and 12122302,
and the Innovative Research Group Project of
Natural Science Foundation of Hunan Province of China No. 2024JJ1008.
}


\clearpage

\begin{table*}
\footnotesize
\caption{\footnotesize
  JWST/NIRSpec-Detected Emission Intensities
  (in Units of $10^{-3}\erg\s^{-1}\cm^{-2}\sr^{-1}$)
  for the C--H and C--D Stretching Bands
  of the H\,{\sc ii} Region, Atomic PDR,
  DF1, DF2 and DF3 Regions in the Orion Bar.
  }
\label{Table:BandInt}
\begin{tabular}{ccccccccc}
\noalign{\smallskip} \hline  \noalign{\smallskip} \hline \noalign{\smallskip}
\multicolumn{2}{c}{Band}
& H\,{\scriptsize II}	& PDR & DF1 & DF2 & DF3 & Note \\ \hline \noalign{\smallskip}
\multicolumn{2}{l}{3.3$\mum$ aromatic CH}
& 10.58 & 45.27 &	30.43 &	24.09 & 15.50	& Peeters et al.\ (2024)	\\ \hline \noalign{\smallskip}
\multirow{2}{8.5em}{4.35$\mum$ \\ aromatic CD, CN$^{a}$}	 &	``generous'' continuum	&	0.98 	            &	5.36 	&	3.78 	&	3.10 	&	2.23 	&	This work	\\ \cline{2-8} \noalign{\smallskip}
	                                                 &	``tight'' continuum	    &	0.71 	            &	--      &	--      &	1.10 	&	0.96 	&	This work	\\ \hline \noalign{\smallskip}
\multirow{2}{8.5em}{4.65$\mum$ \\aliphatic CD$^{b}$}	 &	``generous'' continuum	&	0.60 	            &	1.02 	&	0.92 	&	0.74 	&	0.54 	&	This work	\\ \cline{2-8} \noalign{\smallskip}
	                                                 &	``tight'' continuum	    &	0.40 	            &	--      &		--  &	0.47 	&	0.49 	&	This work	\\ \hline \noalign{\smallskip}
\multirow{2}{8.5em}{4.75$\mum$ \\aliphatic CD$^{c}$}	 &	``generous'' continuum	&	0.35 	            &	0.50 	&	0.51 	&	0.56 	&	0.66 	&	This work	\\ \cline{2-8}\noalign{\smallskip}
	                                                 &	``tight'' continuum	    &	0.25 	            &	--      &		--  &	0.42 	&	0.61 	&	This work	\\ \hline \noalign{\smallskip}
\noalign{\smallskip}
\end{tabular}
a: The band-intensity ratios are
$I_{``4.4''}/I_{3.3}$\,$\approx$\,0.093,
0.118, 0.124, 0.129 and 0.144
for the Orion Bar H\,{\scriptsize II} region,
atomic PDR, DF1, DF2 and DF3, respectively,
if a ``generous'' continuum is adopted
(see Figure~\ref{fig:OrionFit1}).
If we adopt a ``tight'' continuum,
$I_{``4.4''}/I_{3.3}$\,$\approx$\,0.067, 0.046,
and 0.062 for the H\,{\scriptsize II} region,
DF2 and DF3, respectively
(see Figure~\ref{fig:OrionFit2}). \\
b: The band intensity ratios are
$I_{4.65}/I_{3.3}$\,$\approx$\,0.057, 0.023,
0.030, 0.031 and 0.035 for the
H\,{\scriptsize II} region, atomic PDR,
DF1, DF2 and DF3, respectively, if a
``generous'' continuum is adopted.
With a ``tight'' continuum, we obtain
$I_{4.65}/I_{3.3}$\,$\approx$\,0.038, 0.020,
and 0.032 for the H\,{\scriptsize II} region,
DF2 and DF3, respectively.\\
c: The band intensity ratios are
$I_{4.75}/I_{3.3}$\,$\approx$\,0.033, 0.011,
0.017, 0.023 and 0.043 for the H\,{\scriptsize II} region,
atomic PDR, DF1, DF2 and DF3, respectively,
if a ``generous'' continuum is adopted.
By adopting a ``tight'' continuum,
we derive $I_{4.75}/I_{3.3}$\,$\approx$\,0.024,
0.018, and 0.040 for the H\,{\scriptsize II} region,
DF2 and DF3, respectively.
\end{table*}

\clearpage

\clearpage

\clearpage

\clearpage

\clearpage

\clearpage

\clearpage


\begin{thebibliography}{30}
\expandafter\ifx\csname natexlab\endcsname\relax\def\natexlab#1{#1}\fi
%


\bibitem[]{}Allamandola, L.J., Tielens, A.G.G.M., \& Barker, J.R.\
            1985, ApJ, 290, L25

\bibitem[]{}Allamandola, L.~J., Boersma, C.,
                Lee, T.~J., et al.\ 2021,
                ApJL, 917, L35

\bibitem[]{}Bern{\'e}, O., Habart, E., Peeters, E., et al.\
                  2022, PASP, 134, 4301

\bibitem[]{}Boersma, C., Allamandola, L. J., Esposito, V. J., et al.\
                 2023, ApJ, 959, 74

\bibitem[]{}Buragohain, M., Pathak, A., Sarre, P., Onaka, T., \& Sakon, I.\ 2015,
          MNRAS, 454, 193





\bibitem[]{}Chown, R., Sidhu, A., Peeters, E., et al.\
                  2024, A\&A, 685, A75

\bibitem{}Doney, K. D., Candian, A., Mori, T., Onaka, T.,
                \& Tielens, A. G. G. M.\ 2016,
                 A\&A, 586, 65


\bibitem{}Draine, B.~T.\ 2006,
                in Astrophysics in the Far Ultraviolet:
                Five Years of Discovery with FUSE
                (ASP Conf. Ser. 348),
               ed. G. Sonneborn, H. Moos, \& B.-G. Andersson
               (San Francisco, CA: ASP), 58

\bibitem[]{}Draine, B.T., \& Li, A.\ 2001, ApJ, 551, 807

\bibitem[]{}Draine, B.T., \& Li, A.\ 2007, ApJ, 657, 810

\bibitem[]{}Draine B. T., Li A., Hensley B. S., et al.\ 2021, ApJ, 917, 3

\bibitem[]{}Draine, B.T., Sandstrom, K.M., Dale, D.A.,
                  et al.\ 2025, ApJ, submitted



\bibitem[]{}Geballe, T. R., Tielens, A. G. G. M., Allamandola, L. J.,
            Moorhouse, A., \& Brand, P. W. J. L.\
            1989, ApJ, 341, 278

\bibitem[]{}Habart, E., Le Gal, R., Alvarez, C., et al. \
            2023, A\&A, 673, A149

\bibitem[]{}Hudgins, D.~M., Bauschlicher, C.~W., Jr.,
                \& Sandford, S.~A.\ 2004,
                ApJ, 614, 770

\bibitem[]{}Kurucz, R.~L.\ 1979, ApJS, 40, 1


\bibitem[]{}Li, A., \& Draine, B.T.\ 2002, ApJ, 564, 803

\bibitem[]{}Li, A., \& Lunine, J.I.\ 2003, ApJ, 594, 987

\bibitem[]{}Li, K.J., Li, A., Yang, X.J., \& Fang, T.T.\
                 2024, ApJ, 961, 107

\bibitem[]{}Mathis, J. S., Mezger, P. G., \& Panagia, N.\
            1983, A\&A, 128, 212

\bibitem[]{}Menten, K. M., Reid, M. J., Forbrich, J., \& Brunthaler, A.\
                2007, A\&A, 474, 515


\bibitem[]{}Mori, T., Onaka, T., Sakon, I., et al.\ 2022
                ApJ, 933, 35

\bibitem[]{}O'Dell, C. R.\ 2001, ARA\&A, 39, 99

\bibitem[]{}Onaka, T., Mori, T.~I., Sakon, I.,
                et al.\ 2014,
                ApJ, 780, 114

\bibitem[]{}Onaka, T., Sakon, I., \& Shimonishi, T. \ 2022,
                ApJ, 941, 190


\bibitem[]{}Peeters, E., Allamandola, L.~J.,
                 Bauschlicher, C.~W., Jr., et al.\ 2004,
                 ApJ, 604, 252

\bibitem[]{}Peeters, E., Habart, E.,
                 Bern{\'e}, O., et al.\ 2024,
                 A\&A, 685, A74



\bibitem[]{}Sidhu, A., Tielens, A. G. G. M., Peeters, E.,
                 \& Cami, J.\ 2022, MNRAS, 514, 342


\bibitem[]{}Sloan, G.C., Bregman, J.D., Geballe, T.R.,
            Allamandola, L.J., \& Woodward, C.E.\ 1997,
             ApJ, 474, 735

\bibitem[]{}Van De Putte, D., Meshaka, R., Trahin, B., et al.\
          2024, A\&A, 687, A86


\bibitem{}Verstraete, L., Puget, J. L., Falgarone, E., et al.\ 1996,
         A\&A, 315, L337



\bibitem[]{}Yang, X.~J., Glaser, R., Li, A.,
            \& Zhong, J.~X.\
            2016, MNRAS, 462, 1551

\bibitem[]{}Yang, X.~J., Glaser, R., Li, A.,
               \& Zhong, J.~X.\
               2017, New Astron. Rev., 77, 1

\bibitem[]{}Yang, X.~J., Li, A., \& Glaser, R.\
                2020, ApJS, 251, 12




\bibitem[]{}Yang, X.~J., Li, A., He, C.~Y., \& Glaser, R.\
                2021, ApJS, 255, 23


\bibitem[]{}Yang, X.~J., \& Li, A.\
                2023, ApJS, 268, 12


\end{thebibliography}
\end{document}